%% file: main.tex
\documentclass[runningheads]{llncs}
\usepackage{graphicx}
\graphicspath{{gfx/},{experiments/},}  
\usepackage[british]{babel}
\usepackage[T1]{fontenc}
\usepackage[utf8]{inputenc}

\usepackage{lmodern}  
\usepackage{fix-cm}\rmfamily 
\DeclareFontShape{T1}{lmr}{b}{sc}{<->ssub*cmr/bx/sc}{}
\DeclareFontShape{T1}{lmr}{bx}{sc}{<->ssub*cmr/bx/sc}{}
\usepackage[final]{microtype}

\RequirePackage[nolist]{acronym}

\usepackage{wrapfig}

\usepackage{subcaption}

\usepackage{listings,algorithm}
\usepackage[noend]{algpseudocode}
\algdef{SE}[DOWHILE]{Do}{doWhile}{\algorithmicdo}[1]{\algorithmicwhile\ #1}%

\usepackage{amsmath, amssymb, amsfonts, mathtools}

\usepackage[table,dvipsnames]{xcolor}

\usepackage{booktabs}

\usepackage[inline]{enumitem}

\usepackage[nospace]{cite}

\usepackage{xfrac,xspace}
\usepackage{relsize}
\usepackage[normalem]{ulem}

\usepackage{pifont}

\usepackage{tikz}
\usetikzlibrary{positioning}

\usepackage{setspace}

\newcommand{\myparagraph}[1]{\smallskip\noindent \emph{#1}}

\usepackage{comment}
\usepackage{lipsum}

\RequirePackage[%
  bookmarks,
  unicode,
  breaklinks=true,
  ]{hyperref}
  
\usepackage[nameinlink]{cleveref}

\newcommand{\splitatcommas}[1]{%
	\begingroup
	\begingroup\lccode`~=`, \lowercase{\endgroup
		\edef~{\mathchar\the\mathcode`, \penalty0 \noexpand\hspace{0pt plus 1em}}%
	}\mathcode`,="8000 #1%
	\endgroup
}

\makeatletter
\renewcommand{\paragraph}{\@startsection{paragraph}{5}{0em}%
  {.7ex plus .2ex minus .1ex}%
  {-.5em}%
  {\bfseries}}
\makeatother

\makeatletter
\def\orcidID#1{\smash{\href{http://orcid.org/#1}{\protect\raisebox{-1.25pt}{\protect\includegraphics{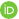}}}}}
\makeatother

\allowdisplaybreaks[4]

\Crefname{equation}{eq.}{eqs.}
\crefname{equation}{equation}{equations}
\Crefname{figure}{Fig.}{Figs.}
\crefname{figure}{figure}{figures}
\Crefname{tabular}{Table}{Tables}
\crefname{tabular}{table}{tables}
\Crefname{definition}{Def.}{Defs.}
\crefname{definition}{definition}{definitions}
\Crefname{proposition}{Prop.}{Props.}
\crefname{proposition}{proposition}{propositions}
\Crefname{section}{Sec.}{Sections}
\crefname{section}{section}{sections}
\Crefname{subsection}{Sec.}{Sections}
\crefname{subsection}{subsection}{subsections}
\crefname{algorithm}{algorithm}{algorithms}
\crefname{listing}{code}{code\ blocks}
\crefformat{footnote}{#2\footnotemark[#1]#3}

\input{defs.tex}

\definecolor{eyecancerpink}{rgb}{1.0, 0.0, 1.0}

\makeatletter%
\g@addto@macro\normalsize{%
  \setlength\abovedisplayskip{3pt}
  \setlength\belowdisplayskip{3pt}%
  \setlength\abovedisplayshortskip{-3pt}%
  \setlength\belowdisplayshortskip{3pt}%
}%
\makeatother

\RequirePackage{graphicx} 
\usepackage[firstpageonly=true,angle=0,vpos=.147\paperheight,hpos=.393\linewidth,vanchor=t,hanchor=l]{draftwatermark} 
\SetWatermarkText{%
\raisebox{-33mm}
{\begin{minipage}{\linewidth}\centering\includegraphics[width=12.5mm]{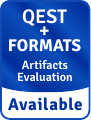}\hfill\includegraphics[width=12.5mm]{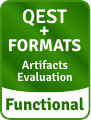}\end{minipage}}%
}

\begin{document}
\title{Time-Sensitive Importance Splitting%
\thanks{%
This work was partially funded by
DFG grant 389792660 as part of \href{https://perspicuous-computing.science}{TRR~248 CPEC},
the European Union (EU) under the INTERREG North Sea project STORM\_SAFE of the European Regional Development Fund,
the EU's Horizon 2020 research and innovation programme under Marie Skłodowska-Curie grant agreement 101008233 (MISSION),
the EU under the Italian National Recovery and Resilience Plan (NRRP) of NextGenerationEU, partnership on ``Telecommunications of the Future'' (PE00000001 program ``RESTART''),
and
by NWO VIDI grant VI.Vidi.223.110 (TruSTy).
}}
\titlerunning{Time-Sensitive Importance Splitting}

%
%
\author{Gabriel~Dengler\inst{1}\orcidID{0000-0002-4217-4952} \and
Carlos E.~Budde\inst{2}\orcidID{0000-0001-8807-1548}\and\newline
Laura~Carnevali\inst{3}\orcidID{0000-0002-5896-4860} \and
Arnd~Hartmanns\inst{4}\orcidID{0000-0003-3268-8674}}
\authorrunning{G. Dengler et al.}
%
\institute{
Saarland University, Saarbrücken, Germany
\\\email{dengler@depend.uni-saarland.de}
\and
Technical University of Denmark, Lyngby, Denmark
\and
Department of Information Engineering, University of Florence, Florence, Italy
\and
University of Twente, Enschede, The Netherlands
}
\maketitle              

\begin{abstract}
State-of-the-art methods for rare event simulation of non-Markovian models face practical or theoretical limits if observing the event of interest requires prior knowledge or information on the timed behavior of the system. In this paper, we attack both limits by extending \acl{ISPLIT} with a time-sensitive \acl{IF}. To this end, we perform backwards reachability search from the target states, considering information about the 
lower and upper bounds of the active timers
in order to steer the generation of paths towards the rare event. We have developed a prototype implementation of the approach for \acl{IOSA} within the \toolset. Preliminary experiments show the potential of the approach in estimating rare event probabilities for an example from reliability engineering. 
\end{abstract}

\acresetall
\setcounter{footnote}{0} 

\renewcommand{\thelstlisting}{\arabic{lstlisting}}  

\input{01-introduction.tex}

\input{02-background.tex}

\input{03-time-sensitive.tex}

\input{04-evaluation.tex}

\input{05-discussion.tex}

\input{06-related-work.tex}

\input{07-conclusion.tex}

\paragraph{Data Availability Statement.}
The artifact of this paper---a reproduction package for the experiments in \Cref{sec:evaluation}---is available at DOI\:\,\href{https://zenodo.org/records/15286766}{\ttfamily\color{blue}10.5281/zenodo.15286766}.

\input{acronyms.tex}

%
%
%
\bibliographystyle{splncs04}
\bibliography{references.bib}

\end{document}

%% file: defs.tex
\newcommand{\N}{\mathbb{N}}

\newcommand{\R}{\mathbb{R}}

\newcommand{\prop}{\ensuremath{\varphi}\xspace}
\newcommand{\est}[1][\prop]{\ensuremath{\hat{#1}}\xspace}

\newcommand{\eminus}[1]{\ensuremath{\scalebox{.85}{\text{\scshape e-}#1}}\xspace}
\newcommand{\cmark}{\ding{51}\xspace}  
\newcommand{\xmark}{\ding{55}\xspace}  
\newcommand{\xtr}[1]{\xrightarrow{\protect{\raisebox{-1pt}[0pt][0pt]{\ensuremath{\scriptstyle{#1}}}}}}

\DeclareMathAlphabet{\mathsc}{OT1}{cmr}{m}{sc}

\newcommand{\dist}[1]{\mathsc{\MakeLowercase{#1}}}
\mathchardef\mhyph="2D 

\let\emptyset\varnothing

\newcommand{\hide}[1]{}

\colorlet{colcodekey}{PineGreen!60!green!70!black}
\colorlet{colcodeidentifier}{Sepia}
\colorlet{colcodenumeric}{blue!25!black}
\colorlet{colcodesyncaction}{violet!90}
\colorlet{colcodecomment}{Black!60}
\colorlet{coldistribution}{Cerulean!60!Black}
\lstdefinelanguage{Kepler}{  
  keywords={},  
  keywords=[2]{toplevel,and,or,pand,vot,fdep,spare,be,sbe,rbox,seq,csp,wsp,hsp},
  keywords=[3]{fail,repair,dorm,prio,fcfs,rand,lambda,prob,res,rate,mean},
  keywords=[4]{exponential,erlang,uniform,normal,lognormal,%
               weibull,gamma,rayleigh,fsteq,dirac,%
               exp,uni,dir},
  otherkeywords={},
  morecomment=**[l]{//},      
  morecomment=**[s]{/*}{*/},
  moredelim=**[is][\color{colcodenumeric}]{^}{^},     
  moredelim=**[is][\color{colcodekey}]{-|}{|-},       
  moredelim=**[is][\slshape]{__}{__},                 
}
\lstdefinestyle{Kepler}{
  language=Kepler,
  xleftmargin=2em,
  basewidth=0.5em,
  basicstyle={\ttfamily},
  identifierstyle=\color{colcodeidentifier},
  keywordstyle=\bfseries,
  keywordstyle=[2]\color{colcodesyncaction},
  keywordstyle=[3]\bfseries\color{colcodekey},
  keywordstyle=[4]\bfseries\color{coldistribution},
  commentstyle=\color{colcodecomment},
  stringstyle=\mdseries,
  showstringspaces=false,
  numbers=left,
  numberstyle=\scriptsize\ttfamily\color{colcodecomment},
  numbersep=0.9em,
  tabsize=4,
  frame=none,
  aboveskip=\bigskipamount,
  belowskip=\medskipamount,
  abovecaptionskip=\smallskipamount,
  belowcaptionskip=\smallskipamount,
  captionpos=b,
  escapeinside={`}{`},
  mathescape=true,  
}
\lstdefinelanguage{IOSA}{
  keywords={},  
  keywords=[2]{dtmc,ctmc,mdp,module,endmodule,properties,endproperties,%
               init,label,formula},
  keywords=[3]{const,global,bool,int,double,clock,P,S,U},
  keywords=[4]{exponential,erlang,uniform,normal,lognormal,weibull,%
               gamma,rayleigh,fsteq},
  otherkeywords={->,:,[,],..,@,\?,\!},
  morecomment=**[l]{//},      
  morecomment=**[s]{/*}{*/},
  morestring=[b]",
  moredelim=**[is][\color{colcodenumeric}]{^}{^},     
  moredelim=**[is][\color{colcodesyncaction}]{~}{~},  
  moredelim=**[is][\slshape]{__}{__},              
}
\lstdefinestyle{IOSA}{
  language=IOSA,
  xleftmargin=\leftmargin,
  basewidth=0.5em,
  basicstyle={\ttfamily},
  identifierstyle=\color{colcodeidentifier},
  keywordstyle=\bfseries,
  keywordstyle=[2]\bfseries,
  keywordstyle=[3]\bfseries\color{colcodekey},
  keywordstyle=[4]\bfseries\color{coldistribution},
  commentstyle=\color{colcodecomment},
  stringstyle=\color{colcodesyncaction},
  showstringspaces=false,
  numbers=left,
  numberstyle=\scriptsize\ttfamily\color{colcodecomment},
  numbersep=0.9em,
  tabsize=4,
  frame=none,
  aboveskip=\bigskipamount,
  belowskip=\medskipamount,
  abovecaptionskip=.5ex,
  belowcaptionskip=\smallskipamount,
  captionpos=b,
  escapeinside={`}{`},
  mathescape=true,  
}

\newcommand{\timers}{\mathcal{T}}
\newcommand{\actions}{\mathcal{A}}
\newcommand{\inactions}{\actions^{\mathsf i}}
\newcommand{\outactions}{\actions^{\mathsf o}}
\newcommand{\coactions}{\actions^{\mathsf u}}
\newcommand{\states}{\mathcal{S}}
\newcommand{\locations}{\mathcal{L}}
\newcommand{\trans}[1][]{\xrightarrow{{#1}}}

\newcommand{\toolset}{\textsc{Modest Toolset}\xspace}

\colorlet{hlboxcoldraw}{black!65}  
\colorlet{hlboxcolfill}{black!6}   
\newsavebox\BODYBOX  
\def\BODY{\unhbox\BODYBOX}
\NewDocumentEnvironment{hlbox}{ O{} }
  {\vspace{1ex plus .5ex minus .3ex}%
	\begingroup\centering\begin{spacing}{1.05}%
	\begin{lrbox}{\BODYBOX}\begin{minipage}{.96\linewidth}%
	\ifthenelse{\equal{#1}{}}{}{\textbf{#1}}}
  {\end{minipage}\end{lrbox}\begin{tikzpicture}%
	\node[rectangle,rounded corners=.3mm,inner sep=1.3ex,thick,
	      draw=hlboxcoldraw,fill=hlboxcolfill] {\BODY};
	\end{tikzpicture}\end{spacing}\endgroup%
	\vspace{1ex plus .5ex minus .3ex}}

%% file: 01-introduction.tex
\section{Introduction}
\label{sec:introduction}

Quantitative evaluation methods provide model-driven guidance for the design, development, and operation of critical systems subject to RAMS requirements~\cite{trivedi2017reliability,LM21,ZZZ21,OK11}.
While the evaluation of Markovian models leverages efficient consolidated approaches~\cite{Gra77,uniformization-algorithm,gspn-paper,BK08}, that of non-Markovian models faces notably harder challenges~\cite{german:1994:supplementary-variables, supplementary-variables-2,vicario2009using}. 
To overcome them, analytical methods provide numerical solutions under modeling restrictions, e.g.,~at most one non-exponential timer in each state~\cite{german:1994:supplementary-variables,telek2001transient,choi1994markov,german2001iterative}, or perform semi-symbolic state space enumeration~\cite{horvath:2012:transient-sscs}, suffering explosion issues with many concurrent non-exponential timers.
Despite recent advances~\cite{amparore2013component,carnevali2021compositional,biagi:2017:exploiting-regenerations,biagi:2017:etcs-level3}, simulation remains the primary method to evaluate large non-Markovian models, as it can limit the computational effort beforehand.

In this context,  notable attention has been paid to \ac{RES}, a method to accelerate crude \ac{MC} simulation 
towards the so-called \emph{rare event}.
One of the two popular \ac{RES} methods 
is \ac{ISPLIT} \cite{LLLT09}, 
which layers (splits) the state space and computes the conditional probabilities of traversing layers until the rare event is reached.
%
There is abundant research on the automatic computation of these levels, tackling complex real-world examples \cite{RBSH13,carlos-budde,BDHS20}. 
Especially compared to the other main \ac{RES} method, \ac{IS}, the precomputations do not require significant amount of human intuition to be made working for a specific example.

Despite this success, \ac{ISPLIT} does not provide the overarching out-of-the-box solution for 
any non-Markovian model,
as it heavily relies on a good 
\ac{IF} to ensure reaching one level from the next does not occur rarely
and thus guarantees progress towards the rare event.
%
Existing techniques, especially \ac{ISPLIT}, fail for a potentially large class of models in which the rare event is caused by rare constellations of the sampled timer values, so that the discrete part of the state alone does not provide enough information to reliably measure the distance to the target~\cite{qest24paper}.
%
%
While the method of~\cite{qest24paper}, using first enumeration of \acp{SSC}~\cite{horvath:2012:transient-sscs} and then \ac{MC} simulation, can theoretically tackle these cases, it only works when the critical event happens near the root of the state-space. 
Furthermore, transient analysis with \acp{SSC} may suffer state space explosion and numerical issues for large and stiff models.
%
%
Potential extensions to address these issues would require a heavy implementation effort, e.g.,~integrating a multilevel switch between \acp{SSC} and simulation whenever encountering a critical event, or exploiting regenerative analysis epochs like in \cite{biagi:2017:exploiting-regenerations}.
%

\paragraph{Contributions.} In this paper, we provide a conceptionally simple yet effective solution to 
these issues
by introducing \emph{time-sensitive \acl{ISPLIT}}.
Compared to conventional \ac{ISPLIT} implementations, it additionally considers lower and upper bounds of the active timers in each state.
%
This allows us to check, e.g., if the target is still reachable, not only given the discrete part of the state but also under consideration of time.
To this end, 
we employ 
an analysis of the timed, but not stochastic, behavior of the system by characterizing the reachability of states (encoding discrete values and timers) by \acp{SC}~\cite{vicario:2001:tpn-analysis}.
As one of our main contributions, we show how classical \ac{SC} analysis for forwards reachability from the initial state can be adapted to compute backwards reachability 
from
the target states.
To the best of our knowledge, our work is the first one to incorporate continuous timer information based on backwards reachability search to provide a
finer-grained
importance metric for \ac{ISPLIT}.
Our approach is attractive for two reasons:

\begin{itemize}
    \item It is \emph{simple}: During the simulation phase, it only requires to 
    replace
    the time-agnostic \ac{IF} with a time-sensitive \ac{IF}. 
    Consequently, we can exploit
    existing research on different \ac{ISPLIT} methods like \ac{FE} \cite{LLLT09} and RESTART \cite{villen-altamirano:1991:RESTART} as well as threshold selection methods \cite{cerou:2007:adaptive,BDH19} 
    with no or only minor modifications (see \Cref{sec:discussion} for the latter). 
    \item 
    It is \emph{effective}, because it provides a 
    finer-grained
    estimation of how important a state is, which is especially apparent if the order of timers is relevant to 
    reach a target state.
    This is an inevitable property in many \ac{DFT} models, especially those with PAND gates, where ordered failures of the attached children 
    must
    happen in order to trigger the PAND's failure.
\end{itemize}
Although our pre-calculations for the time-sensitive \ac{IF} are more computationally expensive than just exploiting
discrete information, especially due to the additional number of 
states to be considered,
the computation times remain manageable for many examples. 
In fact, the considered timing constraints can be efficiently represented by \acp{DBM}~\cite{berthomieu1991modeling,vicario:2001:tpn-analysis}, requiring polynomial time in the number of timers for encoding and manipulation.
%
%
As discussed in \Cref{sec:discussion}, we envision that, in future work, we can achieve both time-sensitive importance estimations and arbitrary scalability by combining our approach with compositional \ac{IF} calculation methods from the literature~\cite{carlos-budde, BDHS20}.

\paragraph{Structure.} 
We first
review necessary background in \Cref{sec:background}, especially on non-deterministic timed analysis with \acp{SC} and \ac{RES}.
After that, we present the main contribution of our paper in \Cref{sec:time-sensitive}, 
whose 
theoretical effectiveness is demonstrated with a simple toy example.
The theoretical claims of our paper are supported by an experimental evaluation in \Cref{sec:evaluation} on a 
more complex
model.
We discuss the practicability of the proposed method in \Cref{sec:discussion}.
Finally, we review related work in \Cref{sec:related-work} and 
draw our conclusions in
\Cref{sec:conclusion}.

%% file: 02-background.tex
\section{Background}
\label{sec:background}


\subsection{Input/Output Stochastic Automata}
\label{sec:background:IOSA}

An \ac{IOSA} with urgency~\cite{DM18} is a tuple $(\splitatcommas{\locations,\actions,\timers,\trans,T_0,l_0})$ consisting of a denumerable set of \emph{locations} $\locations$ with $l_0\in\locations$ the \emph{initial location};
a denumerable set of \emph{actions} $\actions$ partitioned into \emph{input} $\inactions$ and \emph{output} $\outactions$ actions, where $\coactions\subseteq\actions$ are called \emph{urgent};
a finite set of \emph{timers} $\timers$ s.t.\ each $x\in\timers$ has an associated continuous probability measure $\mu_x$ supported on $\R_{\ge0}$, from which $T_0\subseteq\timers$ are called \emph{initial};
and
a transition function
${\trans}\subseteq\locations\times2^\timers\times\actions\times2^\timers\times\locations$
whose elements $\langle l, T, a, T', l' \rangle$ are written $l\xtr{\scriptscriptstyle T,a,T'}l'$.

Elements from $\locations$ are called \guillemotleft states\guillemotright\ in \cite{DM18}---we call them \emph{locations} to emphasize their independence from timers.
We further call elements from $\timers$ \emph{timers}, rather than \guillemotleft clocks\guillemotright, to align our semantics in \Cref{sec:background:SC} with the literature on \acp{STPN} and \acp{SC}.
Urgent actions enable instantaneous communication in this timed setting: \cite{DM18} gives sufficient conditions for the resulting (compositional) semantics to be weakly deterministic.
In particular, Dirac \acp{PDF} are not allowed, and each location $l\in \mathcal{L}$ has a unique set of enabled timers or \guillemotleft active clocks\guillemotright, which we denote with the set $T_l$.
As a result,
closed \ac{IOSA} are fully stochastic, so e.g.\ reachability properties can be approximated by statistical estimators such as those given by \ac{MC} and \ac{RES}  approaches \cite{DM18}.


\smallskip
We call \emph{state} a tuple $\langle l, \tau\rangle \in \mathcal{L} \times (\mathbb{R}_{\geq 0} \ \dot{\cup}\ \{ \bot \})^{\mathcal{T}} = \states$ that contains, besides the location, a valuation for all active timers, which is instantiated via the \emph{timer value function} $\tau$.
%
%
As previously mentioned, for \ac{IOSA} it is known that for each location $l$ only one combination of active timers is possible%
~\cite{MBD20,DM18}.\footnote{%
Our approach also works in the general case, with an initial forwards analysis to find all sets of enabled timers in target locations, and then backwards analysis as in~\Cref{sec:time-sensitive}
}\!
Thus, for a location $l$, it holds for all associated states $s = \langle l, \tau\rangle$ and timers $t\in \mathcal{T}$ that $\tau(t) = \bot$ iff $t\not\in T_l$, viz., when the timer is not active in the current state, it takes the special value $\bot$ that denotes an inactive timer in a location.
%

In the set $T \subseteq \timers$ of a transition $l\xtr{\scriptscriptstyle T,a,T'}l'$, the transition function semantically describes an external event (if $a$ is an input action; then $T = \emptyset$) or the expiring of a timer (if $a$ is an output action; then $|T| = 1$), which triggers action $a$ and starts new timers $T'$.
%
%
Input and output actions facilitate modeling concurrency: Synchronous compositions that are closed (no input actions left) result in a fully stochastic system, which is the focus of this contribution.
For more details about the semantics, we refer the reader to~\cite{DM18}.
Note that there are also other stochastic concurrency models that are heavily used for reliability assessment---one of the most popular being the \ac{STPN}~\cite{paolieri:2021:oris}---that are in essence models equally characterized through locations plus starting and elapsing timers.

\subsection{Non-Deterministic Analysis with State Classes}
\label{sec:background:SC}
An \ac{SC} collects states having the same location and different values of the active timers~\cite{vicario:2001:tpn-analysis,berthomieu1991modeling}.
An \ac{SC} $\Sigma = \langle l, D \rangle$ consists of a location $l \in \mathcal{L}$ and a joint domain~$D$ for the active timers in~$l$, i.e.,~the timers in $T_l$.
For an \ac{SC} $\Sigma = \langle l, D \rangle$, let $\boldsymbol{\tau}$ be the vector of \acp{RV} containing the active timers in $l$.

\myparagraph{Initial \ac{SC}.} 
The initial \ac{SC} $\Sigma_0 = \langle l_0, D_0 \rangle$ collects the initial location~$l_0$ and timer domain 
$D_0 = \bigtimes_{t_i \in T_{l_0}} [a_i,b_i] \subseteq (\mathbb{Q}_{\geq 0} \cup \{\infty\})^{|T_{l_0}|}$ 
with 
$a_i \in \mathbb{Q}_{\geq 0}$,  
$b_i \in \mathbb{Q}_{\geq 0} \cup \{\infty\}$ 
for all $t_i \in T_{l_0}$, 
i.e.,~$D_0$ is the Cartesian product of the initial domains\footnote{
We consider rational bounds for the initial value of each active timer to guarantee that the number of enumerated \acp{SC} is finite, as proved, e.g., by \cite[Lemma~3.2]{horvath:2012:transient-sscs}.} 
of the active timers, from which their values are independently sampled, these timers being in fact independent~\acp{RV}. 
Thus, $D_0$ has a hyper-rectangular shape.

\myparagraph{Successor \acp{SC}.} 
The successor of an \ac{SC} $\Sigma = \langle l, D\rangle$ 
via a transition~$l\xtr{\scriptscriptstyle T,a,T'}l'$
is an \ac{SC} $\Sigma' = \langle l', D'\rangle$ if executing the transition from location~$l$ and timer vector~$\boldsymbol{\tau}$ supported over $D$ yields
location $l'$ and timer vector $\boldsymbol{\tau}'$ supported over $D'$.
In particular, we derive domain~$D'$ from domain~$D$ through the following steps:

\begin{enumerate}
    \item \emph{Conditioning:} 
    Let $T_l = \{t_1,t_2,\ldots,t_n\}$. In a fully deterministic resolved \ac{IOSA}, the timer expiring first, say $t_1$, is contained in the singleton set $T$.
    %
    We condition $\boldsymbol{\tau} = \langle t_1,t_2,\ldots,t_n \rangle$ on $t_1$ expiring first, obtaining
    $\boldsymbol{\tau}_{\alpha} :=
    \langle
    t_1^\alpha, \ldots, t_n^\alpha
    \rangle
    =
    \boldsymbol{\tau}
    \mid t_1 \leq t_i
    $
    $\forall\,i\in\{2,\ldots,n\}$\footnote{
    As in probability theory, we use the symbol $|$ to denote the conditioning on an event.},     
    with support  
    $D_{\alpha} = D \cap \{\tau(t_1) \leq \tau(t_i) \,\forall\,i\in\{2,\ldots,n\}\}$.
    Note that the conditioning step makes the elements of $\boldsymbol{\tau}_a$ be dependent \acp{RV}, and their joint support $D_{\alpha}$ be a non-hyper-rectangular domain. 
    
    %
    
    \item \emph{Time advancement:}
    We reduce the active timers by $t_1^\alpha$ and we drop $t_1^\alpha$ from $\boldsymbol{\tau}_{\alpha}$, 
    yielding 
    $\boldsymbol{\tau}_{\beta} :=
    \langle
    t_2^\beta, \ldots, t_n^\beta
    \rangle
    =
    \langle 
    t_2^\alpha-t_1^\alpha,$ $\ldots,$ $t_n^\alpha-t_1^\alpha 
    \rangle
    $
    with support 
    $D_{\beta}=\{\langle \tau(t_2^\alpha),$ $\ldots,\tau(t_n^\alpha) \rangle \mbox{ s.t. } \exists\,\tau(t_1^\alpha) \mbox{ s.t. }
    \langle \tau(t_1^\alpha), \tau(t_2^\alpha)+\tau(t_1^\alpha),$ $\ldots, \tau(t_n^\alpha)+\tau(t_1^\alpha) \rangle \in D_{\alpha}
    \}$.


    \item \emph{Newly activating:} We add to $\boldsymbol{\tau}_{\beta}$ the $m = |T'|$ timers newly activated in $l'$ (i.e.,~active in $l'$ but not in $l$), say $t_{n+1}, \ldots, t_{n+m}$, which are \acp{RV} independent of $t_{1}^\beta, \ldots, t_n^\beta$. 
    Thus, we obtain 
    $\boldsymbol{\tau}' :=
    \langle
    t_2',\ldots,t_n',t_{n+1}',\ldots,t_m'
    \rangle
    = \langle 
    \splitatcommas{t_{2}^\beta, \ldots, t_n^\beta, t_{n+1}, \ldots, t_{n+m}} 
    \rangle$ 
    with the new support
    $D' = D_{\beta} \times [a_{n+1},b_{n+1}] \times \ldots \times [a_{n+m},b_{n+m}]$.\footnote{In the steps of forwards state space analysis, the subscript~$i$ of each active timer of the domain being computed identifies a timer in the model with bounds $a_i$ and $b_i$.} 
\end{enumerate}
Enumeration of \acp{SC} from the initial \ac{SC} $\Sigma_0$ yields a \ac{SCG} representing the set of timed execution sequences of the IOSA.
Note that---contrary to \acp{TPN} where the \ac{SC} expansion method was originally developed for---an \ac{IOSA} does not support the explicit deactivation of timers when traversing a transition. Each started timer has to elapse to be removed from the set of active timers.
To theoretically incorporate the 
deactivation 
of timers in \ac{SC} analysis, we drop between steps 2 and 3 from $\boldsymbol{\tau}_{\beta}$ the $p$ disabled timers, say~$t_{2}, \ldots, t_{p+1}$, yielding the new timer vector
$\boldsymbol{\tau}_{\gamma} = 
\langle t_{p+2}^{\beta}, \ldots, t_{n}^\beta \rangle$
with support
$\splitatcommas{
D_{\gamma} = 
\{\langle \tau(t_{p+2}^{\beta}),\ldots, \tau(t_{n}^\beta) \rangle 
\mbox{ s.t. } 
\exists\,
\tau(t_{2}^{\beta}),\ldots,\tau(t_{p+1}^{\beta}) \mbox{ s.t. }
\langle 
\tau(t_2^{\beta}),\ldots,\tau(t_{p+1}^{\beta}),\tau(t_{p+2}^{\beta}), \ldots,
\tau(t_n^{\beta}) \rangle \in D_{\beta}
\}
} 
$.
%
%
However, the repairable \acp{DFT} scenarios considered in this paper can be modeled in \ac{IOSA} without deactivation of timers~\cite{MBD20}. Therefore, we are not further considering 
the deactivating step
in the following.

The domain $D$ of each \ac{SC} $\Sigma = \langle l, D \rangle$ turns out to be a \ac{DBM}~\cite{vicario:2001:tpn-analysis,berthomieu1991modeling}, i.e.,~the solution of a set of linear inequalities constraining the differences between pairs of timers (the hyper-rectangular domain~$D_0$ of the initial \ac{SC} $\Sigma_0$ is a special case of \ac{DBM}).
In addition, we use a fictitious timer~$t_{\star}$ to denote the time at which $\Sigma$ is entered, obtaining $D = \{ \tau(t_i) -\tau(t_j) \leq b_{ij}$ $\forall  \, t_i, t_j \in T_{l} \cup\{t_{\star}\} \mbox{ with } t_i \neq t_j\}$.
We represent $D$ by storing its coefficients $b_{ij}$ in a $(|T_{l}| + 1) \times (|T_{l}| + 1)$ matrix, requiring polynomial complexity in the number $|T_{l}|$ of the active timers for manipulation and encoding.
We obtain a unique representation (termed \textit{normal}) of the \ac{DBM} $D$ by interpreting it as a transition matrix of a weighted graph and running the \ac{FW} algorithm to find the shortest paths between all pairs of nodes, which results in the resolution of all transitive implications \cite{vicario:2001:tpn-analysis}.



\subsection{Rare Event Simulation}
\label{sec:background:RES}


\Ac{SMC} can estimate quantitative properties in fully-stochastic formal models, without incurring state space explosion or relying on the memoryless property~\cite{YS02,DLM16}.
For instance, bounded reachability metrics 
boil down to estimating binomial proportions:
Each simulation trace sampled, e.g.\ a sequence $x_i = \{\Sigma_j\}_{j=0}^{n_i}$ of \acp{SC}, either visits a goal location $l_g$ or not.
Then the sample mean for $y_i = 1 \mathop{\mathsf{\underline{if}}} \exists j.\, \Sigma_j = \langle l_g,D \rangle \in x_i \mathop{\mathsf{\underline{else}}} 0 $ is an unbiased estimator of the probability to reach $l_g$ within the given bounds.
For robustness, such point estimates are given with a statistical correctness guarantee, typically a \ac{CI} of width $\pm\epsilon$ that contains the true value $(1-\delta) \cdot 100\,\%$ of the time~\cite{BHMWW25}.
The smaller the $\epsilon$, the more \emph{precise} (and ``better'') the estimation.

When the event of interest occurs with low probability, though, computing precise \acp{CI} needs infeasibly many samples \cite{RT09b}.
\ac{RES} can alleviate this, e.g., by partitioning the state space into layers surrounding the goal states $\states_g\subsetneq\states$, and computing the conditional probabilities of incrementally traversing layers from $\states_0$ to $\states_g$ \cite{Gar00,LLLT09}.
This approach is called \ac{ISPLIT} and its efficiency depends on the way $\states$ is layered.
Such partition is given by an \ac{IF} $f\colon\states\to\N_0$ that maps each state to an \emph{importance} $f(s)\geq0$ s.t.\ $f(s)>f(s')$ suggests a higher probability of observing the rare event when a simulation trace starts from $s$ rather than $s'$.

For formal models and property queries, performant \ac{IF} candidates can be  automatically derived \cite{RBSH13,BDH15,carlos-budde}.
A general approach uses backwards reachability search to compute the number of transitions needed to reach a goal state.
This yields a distance metric $d \colon \states \to \N_0$, and then $f(s) = f(\langle l,\tau\rangle) \doteq \max_{l'\in \mathcal{L}} \{d(l')\} - d(l)$.
While this heuristics has been successfully applied to many case studies \cite{BDH19,BDHS20,BDMS22}, it disregards stochastic information that is not explicit in $\locations$.
In particular, $f$ ignores all timer values $\tau$ that in \ac{IOSA} determine the successor states.
Therefore, the algorithms from \cite{BDH15,carlos-budde} are ineffective to study properties whose low probability depends on the unlikely occurrence of time events, e.g.\ an alarm that fails before a system error, but after the inspection that would have spotted it.

\begin{hlbox}
Our contribution is a novel method to automatically derive a \emph{time-sensitive} importance function that explicitly considers
concrete valuations of (stochastic) timers. 
\end{hlbox}

%% file: 03-time-sensitive.tex
\section{Towards Time-Sensitive Importance Functions}
\label{sec:time-sensitive}

\begin{wrapfigure}[11]{r}{0.36\textwidth}
	\centering
    \vspace{-4ex}
    \includegraphics[width=.82\linewidth]{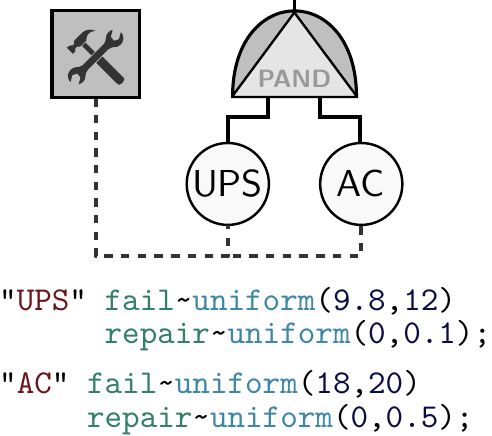}
    \caption{Toy example}
    \label{fig:running-example}
\end{wrapfigure}
We start by studying a small repairable \ac{DFT} example in \Cref{fig:running-example} inspired by \cite{qest24paper}.
It has two \acp{BE}, UPS and AC, that are connected to a PAND gate.
A \ac{TLE}---i.e.\ general system failure---can only happen when UPS fails first and, before it gets repaired, AC fails.
This example can be seen as a simple model of a highly reliable system powered by an unreliable grid, which has a UPS to remain operational during the recurrent blackouts.
The UPS battery is replaced periodically: If during the replacement a blackout occurs, a system failure occurs.
Both \acp{BE} are repaired by a single repair box that can repair one \ac{BE} at a time, and repairs the one which fails first.

\Cref{code:BE_IOSA} shows the \ac{IOSA} \namecref{code:BE_IOSA} for the UPS component.
Its locations are given by the values of variables \lstinline[style=IOSA]{inform} and \lstinline[style=IOSA]{broken}.
Actions \lstinline[style=IOSA]{~down~} and \lstinline[style=IOSA]{~up~} are output, indicated by suffix \lstinline[style=IOSA]{!}, and are broadcast upon the expiration of timers \lstinline[style=IOSA]{fail} and \lstinline[style=IOSA]{repair}, respectively.
These timed events are denoted with \lstinline[style=IOSA]{@} followed by the timer that triggers the transition.
Oppositely, urgent (untimed) actions are taken based on locations alone---\ac{IOSA} operates in a maximal-progress regime.
For example, the \lstinline[style=IOSA]{repair} of \lstinline[style=IOSA]{UPS} does not start immediately after a failure, but rather waits for the RBOX to trigger it via urgent input \lstinline[style=IOSA]{~r~} on line~\ref{code:BE_IOSA:repair}.
This represents a single RBOX that allows one repair at a time, thus implementing interdependent repairs among multiple components.
Full details of these semantics are in \cite{MBD20}.

\begin{lstlisting}[%
    style=IOSA,
    basicstyle=\scriptsize\ttfamily,  % smaller font
    caption=IOSA for the UPS BE in \Cref{fig:DFT}\, following the patterns from \cite{MBD20},
    label=code:BE_IOSA,
    belowskip=-1ex,
    abovecaptionskip=\medskipamount,
    float,
]
module UPS
  fail, repair: clock;
  inform: [^0^..^2^] init ^0^;
  broken: [^0^..^2^] init ^0^;
  // internal semantics for fail and repair, with (timed) output actions
  [~down~!] broken=^0^ @ fail   -> (inform'=^1^) & (broken'=^1^);
  [ ~up~! ] broken=^2^ @ repair -> (inform'=^2^) & (broken'=^0^) & (fail'=uniform(^9.8^,^12^));
  // RBOX triggers our repair, notifying with urgent input action
  [ ~r~?? ] broken=^1^ -> (broken'=^2^) & (repair'=uniform(^0^,^0.1^)); `\label{code:BE_IOSA:repair}`
  // notify state change to parent gates and RBOXes, with urgent output actions
  [~f_i~!!] inform=^1^ -> (inform'=^0^);
  [~r_i~!!] inform=^2^ -> (inform'=^0^);
endmodule
\end{lstlisting}

To demonstrate how one could compute a time-sensitive \ac{IF} in this example, we focus on the shortest path leading from the initial state to the system failure.
Given the supports of the (failure and repair) timer distributions, the \ac{TLE} happens when UPS has failed twice, with a failure time near the lower bound $9.8$, while AC has failed once, with a failure time near the upper bound $20$, and happening after the second failure of UPS.
With this information one can already know in the initial state if the target is reachable on the shortest path. In contrast, using the location information alone to craft an \ac{IF} cannot detect that UPS has to fail \emph{twice} for the shortest path to the target.
In the following, we explain how this information can be determined automatically in a generalized setting.


\paragraph{Timed distance metric.} 
We define the timed distance metric $d : \mathcal{S} \rightarrow \mathbb{N}_0$ of a state $s\in\mathcal{S}$ as the distance of~$s$ to the target given the value of the active timers in~$s$ (rather than as a distance $d : \mathcal{L} \rightarrow \mathbb{N}_0$ based solely on the location of~$s$).
%
%
%
Thus, we obtain as \ac{IF} $f\colon \mathcal{S} \rightarrow \mathbb{N}_0$ with $f(s) = f(\langle l, \tau\rangle) = \max_{s'\in \mathcal{S}} \{d(s')\} - d(s) = \max_{s'\in \mathcal{S}} \{d(s')\} - d(\langle l, \tau\rangle)$.
%
%
%
To calculate our distance metric~$d$, we use the theory of \acp{SC} and effectively run the analysis backwards from the target locations. 

\myparagraph{Target \acp{SC}.} 
For each target location~$l'$, we build the largest possible domain $D'$ for the active timers in~$l'$, considering only whether the target location~$l'$ is reachable from these evaluations of timers, not whether these evaluations of timers are reachable from an initial state.
To this end, we limit each timer $t_i \in T_{l'}$ by its upper bound 
$b_i \in \mathbb{Q}_{\geq 0} \cup \{\infty\}$, obtaining the \ac{SC} $\Sigma' = \langle l', D'\rangle$ with
$D' = \bigtimes_{t_i \in T_{l'}} [0,b_i] \subseteq (\mathbb{Q}_{\geq 0} \cup \{\infty\})^{|T_{l'}|}$.
Conversely, we do not limit $t_i$ by its lower bound $a_i \in \mathbb{Q}_{\geq 0}$ given that, depending on the sequence of executed transitions, it may not be \textit{newly} activated in~$l'$ (i.e., after the last incoming transition).

%
%
%
%
%
%

\myparagraph{Predecessor \ac{SC}.} Given an 
\ac{SC} $\Sigma' = \langle l', D'\rangle$ and an incoming transition $l\xtr{\scriptscriptstyle T,a,T'}l'$, we compute the largest 
possible \ac{SC} $\Sigma = \langle l, D\rangle$ such that for any state $s$ 
in $\Sigma$, the action $a$ leads to a state $s'$ in $\Sigma'$ (we could say that $\Sigma$ is the \emph{weakest precondition} of $\Sigma'$ given~$a$). 
%
%
At each step of this derivation, we normalize the
\ac{DBM} domain of the \ac{SC} being computed using the \ac{FW} algorithm~\cite{vicario:2001:tpn-analysis} (as in 
forwards analysis), ensuring uniqueness of the result and correct resolution of all transitive restrictions. 
In the following derivation steps, let $\boldsymbol{\tau}$ and $\boldsymbol{\tau}'$ be the vectors of \acp{RV} containing the active timers in $l$ and $l'$, respectively, as in \Cref{sec:background:SC}:
%
%
\begin{enumerate}
    \item \emph{Inverse of newly activating:} 
    Let $\boldsymbol{\tau}' := \langle t_2', \ldots, t_{n+m}' \rangle$ and $t_{n+1}', \ldots, t_{n+m}'$ be the $m$ newly activated timers in $l'$ (i.e.,~activated in~$l'$ but not in~$l$).
    We limit each such timer by its lower bound (as no time elapses from newly activation to entrance in $\Sigma'$), obtaining 
    $
    \boldsymbol{\tau}_a :=
    \langle t_2^a, ..., t_n^a, t_{n + 1}^a, ..., t_{n+m}^a \rangle
    =
    \boldsymbol{\tau}' \,\mid\, t_{n+1}', ..., t_{n+m}' \in T' 
    $,
    with support
    $D_a = D' \cap \{ a_i \leq \tau(t_i')\, \forall\, i \in \{n + 1, ..., n+m\} \}$.
    After normalizing $D_a$, we remove 
    $t_{n+1}^a, \ldots, t_{n+m}^a$
    from $\boldsymbol{\tau}_a$, yielding 
    $
    \boldsymbol{\tau}_b := 
    \langle t_2^b, \ldots, t_n^b \rangle =
    \langle t_2^a, \ldots, t_n^a \rangle
    $ 
    with support
    $D_b = \{ \langle \tau(t_2^a), ..., \tau(t_n^a) \rangle$ s.t. $\exists\,\splitatcommas{\tau(t_{n+1}^a), ..., \tau(t_{n+m}^a)}$ s.t. $\langle \tau(t_2^a), ..., \tau(t_n^a), \tau(t_{n+1}^a), ..., \tau(t_{n+m}^a)\rangle \in D_a \}$.

    

    
    \item \emph{Inverse of time advancement:} 
    We increase the timers by the timer expiring first in $\Sigma$, say $t_1^b$, obtaining
    $\splitatcommas{\boldsymbol{\tau}_c
    := 
    \langle 
    t_1^{c}, \ldots, t_n^{c} \rangle
    = 
    \langle t_1^b, t_2^b + t_1^b, \ldots, t_n^b+t_1^b
    \rangle} 
    $ 
    with support 
    $D_c = \{ \langle \tau(t_1^b), \tau(t_2^b), \ldots, \tau(t_n^b) \rangle$ 
    s.t. 
    $\tau(t_1^b) \in [0, \infty)$ 
    s.t. 
    $\langle \tau(t_2^b) - \tau(t_1^b), \ldots, \tau(t_n^b) - \tau(t_1^b)\rangle \in D_b\}$.
    \item \emph{Applying upper bounds of timers:} After normalizing $D_c$, we limit each active timer by its upper bound (to guarantee the timer does not exceed the bound due to the time retardation), obtaining 
    $\boldsymbol{\tau} 
    := \langle t_1, ..., t_n\rangle
    = \boldsymbol{\tau}_c \,\mid\ 
    t_i^c \leq b_i$ 
    $\forall\, i \in \{1, ..., n\}$ 
    with the final domain 
    $D = D_c \cap \{\tau(t_i^c) \leq b_i\, \forall\, i \in \{1, ..., n\}\}$ 
    (which is finally normalized, similarly to the previous passages).

\end{enumerate}
Contrary to forwards analysis,
there is no need for a conditioning step to guarantee that the 
timer $t_1^b$ expiring first
actually elapses
before the other timers. 
In fact, the non-negativity of the timer evaluations in domain $D_b$, 
i.e.,~$0\leq \tau(t^b_i) - \tau(t^b_1)$
$\forall\,i\in\{2, ..., n\}$,
already implies that 
$\tau(t^b_1) \leq \tau(t^b_i)$
$\forall\,i\in\{2, ..., n\}$.\footnote{Note that the conditioning step could also be omitted in forwards analysis when the lower bound of zero is applied to each timer after the time advancement step, as the conditioning step effectively ensures that every timer is non-negative after the time advancement step.}


For each \ac{SC} $\Sigma' = \langle l', D'\rangle$ and incoming transition $l\xtr{\scriptscriptstyle T,a,T'}l'$, we compute the predecessor \ac{SC} $\Sigma = \langle l, D\rangle$ and we derive its distance metric $\omega(\Sigma)$ as the minimum number of transitions needed to reach the target \acp{SC} from $\Sigma$.
During simulation, we evaluate the distance metric of a state~$s=\langle l,\tau \rangle$ as the minimum distance to the target among those of the SCs which $s$ belongs to, i.e., 
$d(s) = \min_{\Sigma \,\mid\, s \in \Sigma} \omega(\Sigma)$. 
A state 
$s=\langle l,\tau \rangle$ 
belongs to an \ac{SC} 
$\Sigma = \langle l',D \rangle$ 
(which we write as $s \in \Sigma$) 
if $l=l'$ and the timer evaluations defined by the timer evaluation function $\tau$ satisfy $D$ (which can be easily checked by evaluating all inequalities encoded in $D$).


\myparagraph{Optimizations.} During calculation of \acp{SC}, we can eliminate redundant information, e.g., if for one location $l$ 
we have two \acp{SC} 
$\Sigma_1 = \langle l, D_1 \rangle $ and
$\Sigma_2 = \langle l, D_2 \rangle $
%
such that $D_1 \supseteq D_2$ 
and
$D_1$ has a lower distance to the target than $D_2$, we can omit 
$\Sigma_2$.
Furthermore, we can eliminate \acp{SC} that have a zero probability to be reached. 
This condition occurs if,
for one non-urgent timer $t_i\in T_l$ (whose support is not a singleton by \ac{IOSA} weakly determinism requirements, 
i.e.,
$a_i < b_i$), the 
timer evaluation
$\tau(t_i)$ is bounded to a single value. 
%
Last but not least,
as we use the distance metric $d(s)$ of a state~$s$ only as a heuristic, 
we can avoid enumerating
the complete \ac{SC} graph; 
we can, e.g., stop after reaching a given expansion depth.

\paragraph{Example derivation.} 
%
In the example of
\Cref{fig:running-example}, we 
have
timers $t_{uf}$ for \emph{\underline{U}PS} 
\emph{\underline{f}ail}, $t_\mathit{ur}$ for \emph{\underline{U}PS \underline{r}epair}, $t_\mathit{af}$ for \emph{\underline{A}C \underline{f}ail}, and $t_\mathit{ar}$ for \emph{\underline{A}C \underline{r}epair} (in the \ac{IOSA} implementation~\cite{MBD20}, we also incorporate urgent actions to coordinate the communication between the two \acp{BE}, the repair box, and the PAND gate; for simplicity, we omit them here).
We start in a target location and traverse the transitions of the \ac{IOSA} backwards, which 
results in inversely following the elapsed timers: 
\begin{itemize}
    \item \emph{Target \ac{SC}:} We consider the first location where the target condition is met, i.e., both components have failed and UPS has failed first. As the repairing of UPS and AC is shared by a repair box, only UPS is actively repaired. Thus, the only active timer is $t_\mathit{ur}$, yielding an \ac{SC} with domain
    $D_1 = \{ 0\leq \tau(t_\mathit{ur})\leq 0.1 \}$.
    \item \emph{Elapsing of $t_\mathit{af}$:} By following backwards the failure of AC, we have as active timers $t_\mathit{ur}$ and $t_\mathit{af}$, obtaining an \ac{SC} with timer domain: 
    \[D_2 = \{ 0\leq \tau(t_\mathit{ur})\leq 0.1\wedge 0\leq \tau(t_\mathit{af}) \leq 0.1\wedge 0\leq \tau(t_\mathit{ur}) - \tau(t_\mathit{af})\leq 0.1\}\]
    Note that the remaining failure time of $t_\mathit{af}$ must be at maximum $0.1$, as otherwise the gate UPS will already be repaired before AC can also fail.
    \item \emph{Elapsing of $t_\mathit{uf}$:} Following the failure of UPS, we reach the initial location, where both the failure timers $t_\mathit{uf}$ and $t_\mathit{af}$ are newly activated:
    \[D_3 = \{0\leq \tau(t_\mathit{uf})\leq 12\wedge0 \leq \tau(t_\mathit{af})\leq 12.1\wedge -0.1\leq \tau(t_\mathit{uf})-\tau(t_\mathit{af})\leq 0 \}\]
    Note that, even though we have reached the initial location, 
    we have not reached an initial state, as the remaining value of the timer $t_\mathit{af}$ must be lower than $12.1$, which is never the case when the timer is newly sampled. Thus, we need to execute further steps in our backwards enumeration of \acp{SC}.
    \item \emph{Elapsing of $t_\mathit{ur}$:} We assume that UPS was repaired again, obtaining domain:
    \begin{align*}
        D_4 &= \{ 0\leq \tau(t_\mathit{ur}) \leq 0.1 \wedge 9.8\leq \tau(t_\mathit{af})\leq 12.2\ \wedge \\
        &\strut\qquad -12.1\leq \tau(t_\mathit{ur}) - \tau(t_\mathit{af})\leq -9.8\}
    \end{align*}
    \item \emph{Elapsing of $t_\mathit{uf}$:} We reach the initial location again,
    with domain:
    \begin{align*}
        D_5 &= \{0\leq \tau(t_\mathit{uf})\leq 10.2\wedge 9.8\leq \tau(t_\mathit{af})\leq 20\ \wedge \\
        &\strut\qquad -12.2\leq \tau(t_\mathit{uf})-\tau(t_\mathit{af})\leq -9.8 \}
    \end{align*}
     The condition $\tau(t_\mathit{uf})-\tau(t_\mathit{af})\leq -9.8 \iff \tau(t_\mathit{uf})+ 9.8\leq \tau(t_\mathit{af})$ is of key importance. In fact, it expresses that, in the time difference between $t_\mathit{uf}$ and $t_\mathit{af}$, an additional failure and repair of the UPS \ac{BE} must take place, which takes at minimum $9.8$ time units in total.
\end{itemize}
The resulting \ac{SC} graph is obviously much larger, as multiple paths are possible to reach the target. However, we depicted this path to underline the effectiveness of how the \ac{ISPLIT} technique can benefit substantially from this information.

%% file: 04-evaluation.tex
\section{Experimental Evaluation}
\label{sec:evaluation}

We use our approach to estimate a (rare) total-failure probability that depends on an ordered sequence of failures and repairs of interdependent components.

\begin{figure}
  \centering
  \begin{minipage}[b]{.33\linewidth}
	\centering
    \includegraphics[width=.8\linewidth]{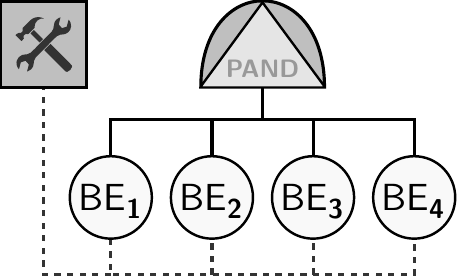}
    \caption{\ac{DFT} with sequential failures and repairs}
    \label{fig:DFT}
  \end{minipage}
  \hfill
  \begin{minipage}[b]{.63\linewidth}
    \def\dist{\raisebox{-.7ex}{\textasciitilde}}
    \begin{lstlisting}[%
        style=Kepler,
		basicstyle=\scriptsize\ttfamily,  % smaller font
        caption=Kepler syntax for the \ac{DFT} in \Cref{fig:DFT}\cite{BDMS22},
        label=code:DFT,
        belowskip=-1ex,
        abovecaptionskip=\medskipamount,
        %float,
    ]
toplevel "PAND1"; 
"PAND1" pand "BE1" "PAND2";
"PAND2" pand "BE2" "PAND3";
"PAND3" pand "BE3" "BE4";
"BE1" fail`\dist`uniform(^1198^,^1218^) repair`\dist`uniform(^10^,^15^);
"BE2" fail`\dist`uniform(^530^,^595^)   repair`\dist`uniform(^10^,^45^);
"BE3" fail`\dist`uniform(^385^,^465^)   repair`\dist`uniform(^10^,^45^);
"BE4" fail`\dist`uniform(^1105^,^1205^) repair`\dist`uniform(^10^,^15^);
"RBOX" rbox prio "BE1" "BE2" "BE3" "BE4";
    \end{lstlisting}
  \end{minipage}
\end{figure}

\subsection{Model and Property}
\label{sec:evaluation:model}

The system under study is depicted in \Cref{fig:DFT}: a synthetic repairable \ac{DFT} whose \ac{TLE} requires the sequential failure of components \lstinline[style=Kepler]{BE1} through \lstinline[style=Kepler]{BE4}.
Their repair and failure distributions make the system failure---i.e.\ the \ac{TLE}---dependent on one failure and no repairs of \lstinline[style=Kepler]{BE1} and \lstinline[style=Kepler]{BE4}, two failures and one repair of \lstinline[style=Kepler]{BE2}, and three failures and two repairs of \lstinline[style=Kepler]{BE4}.

\begin{wrapfigure}[16]{r}{.52\linewidth}
	\centering
	\vspace{-4.5ex}
	\includegraphics[width=\linewidth]{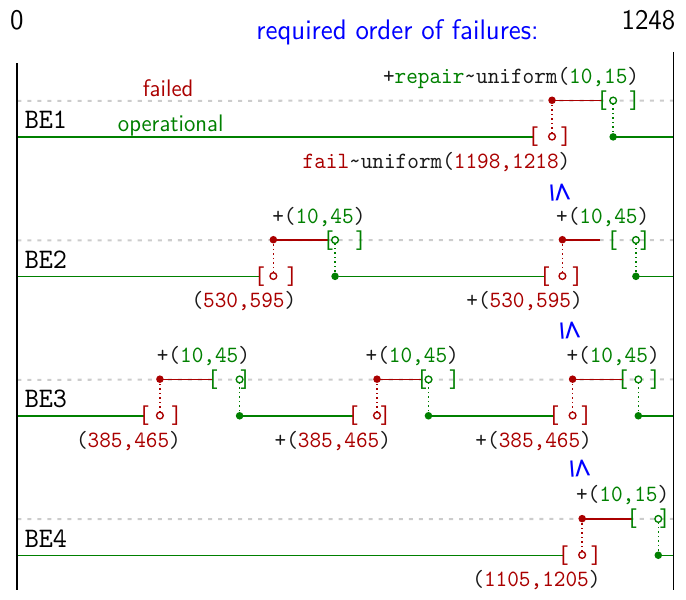}
    \vspace{-3.5ex}
    \caption{Behavior of \ac{DFT} in \Cref{code:DFT}}
    \label{fig:DFT_behaviour}
\end{wrapfigure}
\Cref{fig:DFT_behaviour} depicts the possible order of these events based on the support of their corresponding uniform distributions.
Since the top gate is a PAND gate, a \ac{TLE} occurs when the time
\def\antesque{\raisebox{.25ex}{\ensuremath{\color{blue}\mathsmaller{\boldsymbol\leq}}}}%
 to the last failure of each of these components follows the sequence \lstinline[style=Kepler]{BE1 $\antesque$ BE2 $\antesque$ BE3 $\antesque$ BE4}.

The semantics of this non-Markovian system becomes fully stochastic when given in terms of an \ac{IOSA}~\cite{MBD20}.
We study the transient property of observing a system failure before 1248 time units: $\prop = \Pr(\ac{TLE}\mid t_\textit{age} \leq 1248)$, where $t_\textit{age}$ denotes the global system time.
Given the absence of non-spurious nondeterminism, estimates \est for this property can be computed via \ac{SMC}.

\subsection{Experimental Support in the \texorpdfstring{\toolset}{Modest Toolset}}
\label{sec:evaluation:tool}

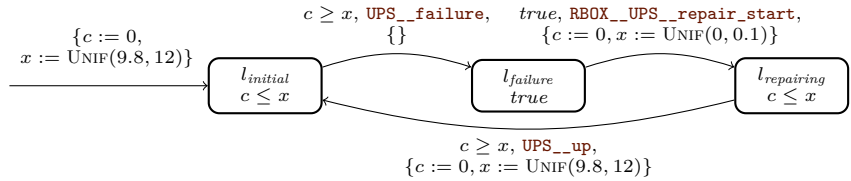
\begin{figure}[t]
    \centering
    \input{gfx/sta_ups_be.tex}
    \caption{Created \ac{STA} for the UPS \ac{BE} in \Cref{fig:running-example}/\Cref{code:BE_IOSA}}
    \label{fig:STA_UPS_BE}
\end{figure}

We have extended the \toolset \cite{HH14} to parse repairable \ac{DFT} models written in Kepler, and compute time-sensitive \acp{IF} from models with stochastic timed semantics.
With the exception of SPARE-gates and -\acp{BE}, the new support for repairable \acp{DFT} covers all gates from \cite{MBD20,BDMS22}, 
including dynamic gates such as PAND (\lstinline[style=Kepler]{pand} in Kepler) and functional dependence (\lstinline[style=Kepler]{fdep}), as well as priority-ordered repair boxes (\lstinline[style=Kepler]{rbox prio}).
The constructed model analyzed by the \toolset is a parallel composition of multiple \acp{STA}~\cite{dargenio-stochastic-automata, stochastic-timed-automata-bbb}.
We mapped the semantics of \ac{IOSA} into the corresponding subset of \ac{STA} by following the patterns from \cite{MBD20}.
As \acp{STA} do not provide explicit support for timers, we implement a timer by using a pair of a clock $c$ and a real-valued random sample $x$, reconstructing the timer value as the difference $x - c$.
%
%
In \Cref{fig:STA_UPS_BE}, we depict the \ac{STA} created 
 for the UPS \ac{BE} from \Cref{fig:running-example}. While \texttt{\color{Sepia}UPS\_\_failure} and \texttt{\color{Sepia}UPS\_\_up} correspond to output actions of the internal failure and repair timer, \texttt{\color{Sepia}RBOX\_\_UPS\_\_repair\_start} is an input action issued by the repair box.

As already mentioned in \Cref{sec:time-sensitive}, we avoid enumerating the complete \ac{SC} graph of the model, which would require significant runtime even for bounded (but high) expansion depths, and which is not required given that we use the \ac{IF} solely as a heuristic to steer the simulation towards the rare event.
%

Furthermore, in the example in \Cref{fig:DFT}, there are only 7 failures and 3 repairs needed to trigger the system failure (see \Cref{fig:DFT_behaviour}), making expansion depths 
below
10 not contributing more significant information.
Still, to assess the impact of the expansion depth, we conduct an ablation study with varying expansion depth.


\subsection{Experimentation and Results}
\label{sec:evaluation:results}

We perform two types of experiments: bounding by execution runtime and by number of simulation runs, in both cases to estimate the (time-)bounded reachability property from \Cref{sec:evaluation:model}.
We estimate the ground truth $\prop\approx5.24\eminus{7}\in[4.4\eminus{7},6.0\eminus{7}]$ by running $318410260 > 2^{28}$ independent \ac{MC} runs (30\,h of runtime).
All experiments were performed on an AMD Ryzen 9 7950X3D CPU with 16 cores running Ubuntu 22.04.
All \ac{RES} simulations used the Fixed Effort \ac{ISPLIT} method with an effort of 16 runs per layer~\cite{Gar00,BDHS20}.

\paragraph{Fixed runtime.}
The first set of experiments fixes an execution wall-clock runtime of 30\:min, and compares the normal approximation \ac{CI} produced by different simulation approaches:
\begin{itemize*}[label=,itemjoin={{; }},itemjoin*={{; and }}]
\item[CMC:]         crude \ac{MC}
\item[RES-notime:]  \ac{ISPLIT} using the time-\emph{agnostic} \ac{IF} from \cite{BDH15,carlos-budde}
\item[RES-time-$d$:]  \ac{ISPLIT} using our new time-\emph{sensitive} \ac{IF} from \Cref{sec:time-sensitive} for various exploration depths~$d$
\end{itemize*}.
\Cref{tab:experiments:runtime} shows the results of these experiments, where the second and third columns show the computed estimate \est and the width of the corresponding \ac{CI} (i.e.\ $\est\pm\epsilon$), respectively; \#runs is the number of simulation runs achieved in the given runtime; and \prop? (resp.\ 0?) indicates whether $\prop\in\est\pm\epsilon$ (resp.\ $0\in\est\pm\epsilon$).

Generally speaking for statistical results, the smaller the $\epsilon$ the better the estimation.
In \Cref{tab:experiments:runtime} the best method in this respect was \ac{ISPLIT} using the time-sensitive \ac{IF} for an exploration depth of 10 (shadowed row).
Also, all RES-time-$d$ runs except $d=22$ performed better than both CMC and RES-notime.
Further quality criteria are whether the \ac{CI} contains \prop (good\;\cmark), and whether it contains 0 (bad\;\xmark, because $\prop>0$): All runs using a time-sensitive \ac{IF} satisfied these criteria, again with the exception of RES-time-22.

The under-performance of \ac{IF} for exploration depths over 20 is caused by the low number of runs achieved, in turn caused by the simulation overhead derived from a splitting of the state space beyond the region of interest.
As discussed in \Cref{sec:evaluation:tool}, the \ac{TLE} occurs after 10 timer events, so exploration depths beyond 10 create layers in $\states$ that are irrelevant for \prop.
As a result, executing \ac{FE} in those layers uses simulation budget that does not result in a reduced variance of the estimator.
Such issues are typically alleviated by selecting \emph{thresholds}: a subset of the \ac{IF} values that the \ac{ISPLIT} algorithms use for simulation.
However, this does not help when the rare condition is to choose a few selected paths among many, all of the same length.
This is the case for time-bounded properties, like in our example, which would instead benefit from adding the global time to the information used by the \ac{IF}.
We comment on these aspects in \Cref{sec:discussion}.

\begin{table}[t]
    \centering
	\caption{Results of experimentation on the \ac{DFT} from \Cref{code:DFT}}
	\label{tab:experiments}
	\vspace{-1ex}
	\begin{subtable}{0.48\linewidth}
	\raggedright
	\caption{Fixed execution runtime: 30~min}
	\label{tab:experiments:runtime}
    \smaller
	\begin{tabular}{    l@{~}
	                >{~}c@{~}
	                >{~}c
	                    r
	                >{~}c
	                    c}
		\toprule
		\bfseries Method
		& $\boldsymbol{\est}$
		& $\boldsymbol{\epsilon}$
        & \multicolumn{1}{c}{\#runs}
		& $\boldsymbol{\prop}$?
		& $\boldsymbol{0}$? \\
		\midrule
		CMC         & $1.8\eminus{7}$ & $3.6\eminus{7}$ & 5445630\! & \cmark & \xmark \\
		RES-notime  & $7.1\eminus{7}$ & $7.0\eminus{7}$ & 320736 & \cmark & \\
		RES-time-6  & $5.2\eminus{7}$ & $2.0\eminus{7}$ & 332912 & \cmark & \\
		\rowcolor{Black!9}
		RES-time-10 & $4.3\eminus{7}$ & $1.4\eminus{7}$ & 334507 & \cmark & \\
		RES-time-14 & $5.3\eminus{7}$ & $1.8\eminus{7}$ & 316633 & \cmark & \\
		RES-time-18 & $5.3\eminus{7}$ & $1.8\eminus{7}$ & 227386 & \cmark & \\
		RES-time-22 & $1.1\eminus{6}$ & $4.6\eminus{7}$ & 151576 &        & \\
		\bottomrule
	\end{tabular}
	\end{subtable}
	\hfill
	\begin{subtable}{.48\linewidth}
    \raggedleft
	\caption{Fixed number of simulations: 50000}
	\label{tab:experiments:numruns}
    \smaller
	\begin{tabular}{    l@{~}
					>{~}c@{~}
				    >{~}c
				    >{~}r
				    >{~}c    
				        c}
		\toprule
		\bfseries Method
		& $\boldsymbol{\est}$
		& $\boldsymbol{\epsilon}$
		& \multicolumn{1}{c}{time}
		& $\boldsymbol{\prop}$?
		& $\boldsymbol{0}$? \\
		\midrule
		CMC         & 0               & 0               &  16.4\,s &        & \xmark \\
		RES-notime  & 0               & 0               & 277.4\,s &        & \xmark \\
		RES-time-6  & $3.6\eminus{7}$ & $4.6\eminus{7}$ & 266.0\,s & \cmark & \xmark \\
		\rowcolor{Black!9}
		RES-time-10 & $3.3\eminus{7}$ & $2.6\eminus{7}$ & 270.7\,s & \cmark & \\
		RES-time-14 & $6.4\eminus{7}$ & $5.3\eminus{7}$ & 285.1\,s & \cmark & \\
		RES-time-18 & $8.7\eminus{7}$ & $6.2\eminus{7}$ & 403.3\,s & \cmark & \\
		RES-time-22 & $4.7\eminus{7}$ & $4.3\eminus{7}$ & 583.1\,s & \cmark & \\
		\bottomrule
	\end{tabular}
	\end{subtable}
\end{table}

\paragraph{Fixed number of runs.}
The second set of experiments performs 50000 simulation runs per method, to see which \ac{IF} deploys the most efficient \ac{RES} under a very limited simulation budget, that also avoids the \#runs imbalance observed in \Cref{tab:experiments:runtime}.
The simulation budget chosen is tight, to implement a stress test that evaluates how well the time-sensitive \ac{IF} can observe rare events in a low number of trials.
In this setting, the execution runtime becomes a relevant dependent variable: the faster the better.
Results for experimentation with the same \ac{ISPLIT} configurations as before are shown in \Cref{tab:experiments:numruns}.

Here, CMC and RES-notime fail to reach the rare event, but the latter takes longer because the effort per simulation is higher in \ac{FE} RES than in CMC.
In contrast, all RES-time-$d$ cases managed to produce useful results, in runtimes comparable (sometimes faster) than RES-notime.
The best run was again obtained with RES-time-10.
However, and in contrast to the previous results, here all RES-time-$d$ runs produced \acp{CI} that contain the ground truth \prop.
We highlight that the simulation budget chosen is low enough to make even the time-sensitive \ac{IF} produce bad runs on occasion.
This is a real stress test to explore the boundaries of our implementation---e.g.\ for 20k runs the results were too unstable, while for 50k runs they appeared to be sufficiently stable.
%
%

%% file: gfx/sta_ups_be.tex
\begin{tikzpicture}[font=\scriptsize]
    \node[draw, thick, rounded corners, minimum width=1.5cm, align=center] (A) at (0, 0) {$l_\textit{initial}$\\ $c \leq x$};
    \node[draw, thick, rounded corners, minimum width=1.5cm, align=center] (B) at (3.5, 0) {$l_\textit{failure}$\\ $\textit{true}$};
    \node[draw, thick, rounded corners, minimum width=1.5cm, align=center] (C) at (7, 0) {$l_\textit{repairing}$\\ $c \leq x$};

    \draw[->] (A) edge[bend left=18] node[above, align=center]{$c\geq x$, \texttt{\color{Sepia}UPS\_\_failure},\\ $\{\}$} (B);
    \draw[->] (B) edge[bend left=18] node[above, align=center]{$\textit{true}$, \texttt{\color{Sepia}RBOX\_\_UPS\_\_repair\_start},\\ $\{c:=0, x := \textsc{Unif}(0,0.1)\}$} (C);
    \draw[->] (C) edge[bend left=14] node[below, align=center]{$c\geq x$, \texttt{\color{Sepia}UPS\_\_up},\\ $\{c:=0, x := \textsc{Unif}(9.8,12)\}$} (A);
    \draw (-3.4, 0) edge[->] (A);
    \node[align=center] at (-2.1, 0.5) {$\{c:=0,$\\ $x:= \textsc{Unif}(9.8, 12)\}$};
\end{tikzpicture}

%% file: 05-discussion.tex
\section{Discussion}
\label{sec:discussion}

The experiments in \Cref{sec:evaluation} show how our time-sensitive \acp{IF} can help in practical scenarios to significantly increase the precision of estimates for time-bounded reachability properties under different simulation time budgets.
However, our model was specifically designed for this aim, exacerbating different distances to the target (from the initial location) for different timer values.
Moreover and albeit non-trivial, the model is quite small and rigid, e.g.\ displaying only uniform distributions.
In this section, we discuss the most relevant limitations of our approach in its current implementation, devising posible ideas to tackle them.


\myparagraph{State space explosion.} In the given example, we limited the expansion depth to reduce the effects of the state space explosion. 
%
This issue is especially exacerbated
when the bounds of the 
supports of the active timers
contain many decimal places, as then they likely have a small common factor, leading to many resulting \acp{SC}. 
As 
discussed
in \Cref{sec:time-sensitive}, 
we can overcome this problem by resorting to partial \ac{SCG} enumeration.
%
However, in larger examples, the expansion depth might be too small for a meaningful \ac{IF}. Ideas to mitigate this problem are:

\begin{itemize}
    \item \emph{Compositional \ac{IF} calculation methods:} As the state space explosion problem also occurs in time-agnostic backwards search methods when considering large enough concurrent models, researchers have developed methods to extract reachability information from local components independently and combine the local information of each component into one compositional \ac{IF} \cite{carlos-budde}, with applications to tandem queues, database systems, and oil pipelines. For recursive models like \acp{DFT}, the \ac{IF} can also be computed in a recursive manner~\cite{BDHS20}. Similar techniques could also be applied to our setting. Contrary to the time-agnostic setting, the interesting information is usually detected when observing synchronizations of concurrent components (see the models in \Cref{sec:time-sensitive} and \Cref{sec:evaluation}), making it natural to compute local importance functions based on the synchronizations of a few local components. It remains an open challenge to automatically detect which components should be synchronized.
    
    \item \emph{Improved duplication detection methods:} As described in \Cref{sec:time-sensitive}, for every \ac{SC} $\langle l, D_1\rangle$, we check if there is another \ac{SC} $\langle l, D_2\rangle$ with the same location $l$ and lower distance to the target such that $D_1 \subseteq D_2$, and we remove $\langle l, D_1\rangle$ if this 
    condition occurs.
    %
    Thus, we are only checking for the total inclusion of one \ac{DBM} zone into another. 
    However, it could also happen that there exists a set of \acp{SC}  $\langle l, D_2\rangle$, \ldots, $\langle l, D_k\rangle$ (with $k > 2$) with lower distance to the target than $D_1$ such that $D_1 \subseteq D_2 \cup \dots \cup D_k$. Also in this case, domain $D_1$ becomes redundant and can be discarded, although it is not fully contained in any of the domains $D_2$ to $D_k$. Given that \ac{DBM} zones are not closed under union operations, providing an efficient implementation is challenging.
    \item \emph{Neglecting distribution bounds:} 
    %
    We can reduce the state space size
    by neglecting the
    bounds 
    of timers
    and only considering the order in which the timers 
    have to elapse 
    to reach the target states. 
    We can implement this approach by considering the bounds $[0, \infty)$ for each timer and running backwards analysis with \acp{SC} as described in \Cref{sec:time-sensitive}.
    %
    %
    Although this solution is ineffective in many scenarios, including the 
    example in \Cref{fig:running-example}, the approach can still be effective in timer value order-relevant decision-making with a significantly larger state space (e.g., for repairable \ac{DFT} models featuring many PAND gates).
\end{itemize}

\myparagraph{Timers with infinite support.} In the current experiments, we only 
consider
variables that follow the uniform distribution and thus have lower and upper bounds. Many distributions used in real-world scenarios have unbounded support, with the exponential distribution
a prominent example. We could resort to the actual bound $[0, \infty)$ in the backwards \ac{SC} calculations, 
though
the efficiency of the time-sensitive \ac{IF} lives from providing tight bounds for each 
timer.

To handle this problem, we observe that distributions with infinite support mostly sample on a small finite band. For demonstration, consider a random variable $X$ that is exponentially distributed with rate $1$. We can easily calculate that $\Pr(X \geq x) = 1 - P(x \leq X) = 1 - (1 - e^{-x}) = e^{-x}$, which means that the probability is exponentially decaying for larger selections of $x$. Given that already $Pr(X \geq 10) \approx 4.5\cdot 10^{-5}$, we can say that almost certainly sampled values for $X$ are in the domain $[0, 10]$. Moreover, the confidence of this statement can be strengthened by further increasing the value of $x$.

The above consideration does not account for cases where the rare event depends on heavy tails, or even e.g.\ sampling large values from exponential distributions.
These cases cannot be reliably solved by \ac{ISPLIT} in models where the simulation granularity stops at the level of sampling values from continuous \acp{RV}.
Instead, such problems can be tackled by modifying the distributions of the \ac{RV} as in \ac{IS} \ac{RES}~\cite{LMT09,RBSJ18}, 
or using analytical expansion with \acp{SSC}~\cite{qest24paper}.

\myparagraph{Time until simulation end.} The current backwards reachability approach does not incorporate the elapsed global time and thus ignores the time until the end of the simulation in a time-bounded setting, like the one studied in our work.
Especially when the remaining simulation time is low, it could be useful to know whether the target can still be reached given the time bound.
For example, consider the 
example in \Cref{fig:running-example} and assume that we want to reach the target from the initial location in 21 time units.
We can achieve this objective only following
%
%
the direct path to the target featuring two failures of UPS and then the failure of AC, as any other path would take longer and the target would not be reachable in time.
To this end, we can incorporate
%
%
a global timer $t_\textit{age}$ (with possible negative values) that is never removed from the computed \acp{SC}~\cite{horvath:2012:transient-sscs} and allows us to extract the minimum and maximum time needed until reaching the target.
This solution has also been approached via \emph{forcing} in \ac{IS}, where the next firing values are sampled conditional to being smaller than the time bound \cite{NSN01,BRS20}.

\myparagraph{Automatic threshold selection.}
An important precomputation step to yield performant \ac{ISPLIT} implementations is threshold selection, viz.\ determining a subset of the importance values that will effectively be used to partition the state space.
On the one hand, thresholds chosen too close together
may incur a splitting overhead, and ultimately a runtime explosion.
On the other hand, thresholds chosen too far away from each other can lead to an extinction phenomenon, e.g., no simulation path gets from one threshold to another~\cite{carlos-budde,BDH19}.

Our experiments in \Cref{sec:evaluation} set a threshold for every \ac{IF} value.
As the computation effort using \ac{FE} as \ac{ISPLIT} method is limited by design, using all levels as thresholds is acceptable for a preliminary evaluation---in this sense, \ac{FE} with static width $1$ is also used as basis for the expected success threshold selection method \cite{BDH19}.
Still, to increase the performance of \ac{ISPLIT} techniques, the threshold-selection phase should be performed after the \ac{IF} construction phase.

One challenge in this regard is that threshold selection methods start in the state with lowest importance.
In most models, this condition simply applies to the initial states. However, in timed settings, different states belonging to the same location could have a different importance value (as demonstrated in \Cref{sec:time-sensitive}).
A possible way to tackle this problem is applying rejection sampling to determine states in the initial location with the lowest importance, and bootstrap the threshold selection algorithms from those states.

\myparagraph{Confidence intervals for rare events.}
One of the metrics used to compare the quality of the \acp{IF} in \Cref{sec:evaluation:results} is the width of the resulting \ac{CI}.
However, these were normal (central limit theorem) intervals, which are known to be unsound~\cite{BHMWW25}.
The problem with using sound intervals such as those based on the \ac{DKW} inequality is that they are largely uninformative of the quality of the rare event estimation, precisely by virtue of them providing safe bounds for any sample.
In particular, \ac{DKW} intervals were practically the same for cases like CMC and RES-time-10 (and -6) in \Cref{tab:experiments:numruns}, even though 
RES-time-10
provides an estimate (and the normal \ac{CI} contains \prop), while 
CMC
observed no rare event whatsoever.
To the best of our knowledge, no sound \ac{CI} provides relevant information in rare event regimes, hence we used the (unsound) standard intervals.
We consider this an open problem worthy of investigation.

\myparagraph{\texorpdfstring{\ac{RES}}{RES} algorithm.}
We performed our experiments using a single \ac{ISPLIT} simulation algorithm, namely \ac{FE} with 16 runs per layer.
While a more comprehensive experimentation would be beneficial, the results would only be relevant in comparisons to CMC, which is anyway known to fail for extreme rare regimes.
The main goal of our experimentation was to assess the quality of the (automatic) time-sensitive \ac{IF} when compared against time-agnostic alternatives, for which the use of different \ac{ISPLIT} simulation algorithms is only marginally relevant \cite{BDH19}.




%% file: 06-related-work.tex
\section{Related Work}
\label{sec:related-work}


Niehage et al.~\cite{niehage:2025:isplit-hybrid-systems} recently defined a method to automatically obtain an \ac{IF} for \acp{HPnG} based on the \ac{PLT}~\cite{huels:2021:parametric-location-tree-1, niehage:2025:parametric-location-tree-2}, which is a symbolic representation of the embedded non-stochastic model's state space. 
Like \acp{SC}, each parametric location represents an uncountable set of states of the \ac{HPnG}; they are connected via discrete events (e.g., firing of a transition).
Similar to our work, the \ac{IF} then considers the distance of each symbolic location to the target.
However, where we \emph{conceptually} start from the target states, \cite{niehage:2025:isplit-hybrid-systems} first performs a forwards expansion and then uses the information gained to search backwards in the computed \ac{PLT}.
This solution is less informative than our method, as it does not reflect that different parts of a node in the \ac{PLT} could have a different distance to the target, e.g., the initial location, as we demonstrated in \Cref{sec:time-sensitive}.

Our work is also related to the zone-based backwards reachability analysis for \acp{TA} and \acp{PTA}~\cite{bouyer:2011:ta-backwards-reachability}.
Zones in \acp{TA} and \acp{PTA} are similar to \acp{SC} in the sense that they represent infinitely many configurations considering a location and can be efficiently stored with \acp{DBM}, but they work with models with increasing clocks instead of decreasing timers.
Backwards reachability is, e.g., used in the model checking of \ac{TCTL} formula against \acp{TA} \cite{henzinger:1992:tctl-model-checking} or \ac{PTCTL} formula against \acp{PTA} \cite{KWIATKOWSKA20071027, backwards-reachability-pta}. 
\ac{SMC} with \ac{RES} for stochastic \ac{TA}-based models is also implemented in \textsc{Uppaal SMC}~\cite{DLLMW11}, using \ac{IS}~\cite{JLLMPS16} with a symbolic forwards exploration to identify states that cannot reach the goal.

We note that we effectively modeled the semantics of repairable \acp{DFT} with \acp{STA} instead of \ac{IOSA}, as that is what the \toolset's simulation engine works with (see \Cref{sec:evaluation:tool}).
We thus defined timer values via pairs of clocks and stochastic variables (e.g., clock $c$ with variable $x := \text{Uniform}(1, 2)$).
One may be tempted to construct an \ac{IF} based on the backwards search of an abstracted \acp{TA} of the original \ac{STA}, where the bounds of the support of a stochastic variable are used in location invariants and transition guards (e.g., turning $c \leq x$ and $c \geq x$ into $c\leq 2$ and $c\geq 1$).
Unfortunately, this would 
not incorporate the actual sampled values for stochastic variables like $x$, which is however crucial to estimating the distance to the target (as illustrated in \Cref{sec:time-sensitive}).

%% file: 07-conclusion.tex
\section{Conclusion}
\label{sec:conclusion}

In this paper, we introduced \emph{time-sensitive \acl{ISPLIT}} as a novel extension of classical \ac{ISPLIT} for the \ac{IF} calculation, which incorporates the timer values in estimating the distance to the target states. We started by studying the necessity for considering timer information in \ac{ISPLIT} techniques for specific scenarios where intricate timer constellations make the location information alone not sufficient to obtain an adequate distance measure to the target states. Based on the existing theory of \acp{SC}, we presented an automatic way to calculate this information in a backwards reachability manner.
We implemented a prototype upon the \toolset.
The empirical evaluation based on a repairable \ac{DFT} example with PAND gates demonstrated very encouraging results.

In future work, we plan to realize some ideas to 
tackle
the shortcomings of the current basic method as discussed in \Cref{sec:discussion}. 
Specifically, our first focus will be on providing an extended evaluation, considering larger and more realistic examples, potentially featuring also  other gate types, e.g., SPARE gates.
Apart from that, it will be a theoretically exciting avenue to study backwards reachability for other model classes, e.g. \acp{STA} or hybrid systems.


%% file: acronyms.tex

\begin{acronym}[ABCDEFGHIJK]
    \acro{AP}{atomic proposition}
    \acro{BE}[BE]{basic event}
    \acro{BFS}[BFS]{breadth-first search}
    \acro{BM}[BM]{Bernstein mixed polynomial and expolynomial}
    \acro{BP}[BP]{Bernstein polynomial}
    \acro{CDF}[CDF]{cumulative distribution function}
    \acro{CI}[CI]{confidence interval}
    \acro{CSL}{continuous stochastic logic}
    \acro{CTMC}[CTMC]{continuous time Markov chain}
    \acro{DET}[DET]{deterministic}
    \acro{DBM}[DBM]{difference bounds matrix}
    \acro{DFT}[DFT]{dynamic fault tree}
    \acro{DKW}[DKW]{Dvoretzky–Kiefer–Wolfowitz}
    \acro{DSPN}[DSPN]{deterministic and stochastic Petri net}
    \acro{DTMC}[DTMC]{discrete time markov chain}
    \acro{EXP}[EXP]{exponential}    
    \acro{ERTMS}[ERTMS]{European Rail Traffic Management System}
    \acro{ES}[ES]{expected success}
    \acro{ETCS}[ETCS]{European Train Control System}
    \acro{FE}[FE]{fixed effort}
    \acro{FS}[FS]{fixed success}
    \acro{FTA}[FTA]{fault tree analysis}
    \acro{FT}[FT]{fault tree}
    \acro{FW}[FW]{Floyd-Warshall}
    \acro{GEN}[GEN]{general}
    \acro{GSPN}[GSPN]{generalized stochastic Petri net}
    \acro{HPnG}[HPnG]{hybrid Petri net with general transitions}
    \acro{IF}[IF]{importance function}
    \acro{IMM}[IMM]{immediate}
    \acro{IOSA}[IOSA]{input/output stochastic automata}
    \acro{IS}[IS]{importance sampling}
    \acro{ISPLIT}[ISPLIT]{importance splitting}
    \acro{JPD}[JPD]{joint probability distribution}
    \acro{MCMC}{Markov chain Monte Carlo}
    \acro{MC}[MC]{Monte Carlo}
    \acro{MH}{Metropolis-Hastings}
    \acro{MRP}[MRP]{Markov regenerative process}
    \acro{MRnP}[MRnP]{Markov renewal process}
    \acro{MSE}[MSE]{mean-squared error}
    \acro{NDA}{non-deterministic analysis}
    \acro{PTA}[PTA]{probabilistic timed automaton}
    \acro{PTCTL}[PTCTL]{probabilistic \ac{TCTL}}
    \acro{PDF}[PDF]{probability density function}
    \acro{PN}[PN]{Petri net}
    \acro{PLT}[PLT]{parametric location tree}
    \acro{RESTART}[RESTART]{repetitive simulation trials after reaching thresholds}
    \acro{RES}[RES]{rare event simulation}
    \acro{SBE}[SBE]{spare basic element}
    \acro{SC}[SC]{state class}
    \acro{SCG}[SCG]{state class graph}
    \acro{SEQ}[SEQ]{sequential Monte Carlo}
    \acro{SMC}[SMC]{statistical model checking}
    \acro{SSC}[SSC]{stochastic state class}
    \acro{STA}[STA]{stochastic timed automaton}
    \acro{STPN}[STPN]{stochastic time Petri net}
    \acro{TA}[TA]{timed automaton}
    \acro{TCTL}[TCTL]{timed computation tree logic}
	\acro{TLE}[TLE]{top-level event}
    \acro{TPN}[TPN]{time Petri net}
    \acro{VI}[VI]{value iteration}
    \acro{i.i.d.}[i.i.d.]{independent and identically distributed}
    \acro{w.l.o.g.}[w.l.o.g.]{without loss of generality}
    \acro{RV}[RV]{Random Variable}
    \acroplural{BE}[BEs]{basic elements}
    \acroplural{BM}[BMs]{Bernstein mixed polynomials and expolynomials}
    \acroplural{BP}[BPs]{Bernstein polynomials}
    \acroplural{CI}[CIs]{confidence intervals}
    \acroplural{CTMC}[CTMCs]{continuous time Markov chains}
    \acroplural{DBM}[DBMs]{difference bound matrices}
    \acroplural{DTMC}[DTMCs]{discrete time Markov chains}
    \acroplural{FT}[FTs]{fault trees}
    \acroplural{GSPN}[GSPNs]{generalized stochastic Petri nets}
    \acroplural{HPnG}[HPnGs]{hybrid Petri nets with general transitions}
    \acroplural{MRP}[MRPs]{Markov regenerative processes}
    \acroplural{MRnP}[MRnPs]{Markov renewal processes}
    \acroplural{PTA}[PTA]{probabilistic timed automata}
    \acroplural{PN}[PNs]{Petri nets}
    \acroplural{RV}[RVs]{random variables}
    \acroplural{SC}[SCs]{state classes}
    \acroplural{SMP}[SMPs]{semi-Markov processes}
    \acroplural{SSC}[SSCs]{stochastic state classes}
    \acroplural{STA}[STA]{stochastic timed automata}
    \acroplural{STPN}[STPNs]{stochastic time Petri nets}
    \acroplural{TA}[TA]{timed automata}
	\acroplural{TLE}[TLEs]{top-level events}
    \acroplural{TPN}[TPNs]{time Petri nets}
\end{acronym}